\documentclass[twoside]{dis04}

\usepackage{cite}

\def\runauthor{S.V.Chekanov}
\def\shorttitle{Results of the
searches for narrow baryonic states in DIS 
}

\newcommand{\ks}{K^{0}_{S}}
\newcommand{\xp}{\Xi^{-} \> \pi^{\pm}}
\newcommand{\xbp}{\bar{\Xi}^{+} \> \pi^{\pm}} 
\newcommand{\ksp}{K^{0}_{S}\> p}
\newcommand{\kspb}{K^{0}_{S}\> \bar{p}}
\newcommand{\ksppb}{K^{0}_{S}\> p\>(\bar{p})}
\newcommand{\coll}{Collaboration}
\newcommand{\etal}{ {\it et al.,} }
\newcommand{\gev}{\; \mathrm{GeV}}
\newcommand{\mev}{\; \mathrm{MeV}}

\begin{document}

%  \hfill {ANL-HEP-CP-04-48}

%Title of paper
\title{Results of the searches for narrow baryonic states with
strangeness  in DIS at HERA}

\author{S.~Chekanov \\ \small{for the ZEUS Collaboration} }

\address{HEP division, Argonne National Laboratory,
9700 S.Cass Avenue, \\
Argonne, IL 60439
USA \\ E-mail: chekanov@mail.desy.de}

\maketitle

\abstracts{Searches for narrow baryonic states in the $\ksp$,
$K^+p$, $\xp$ and $\xbp$  decay channels are reported. The data were 
collected with the ZEUS detector at HERA using an integrated 
luminosity of 121 pb$^{-1}$. The searches were  performed in the
central rapidity region of inclusive deep inelastic scattering at
an $ep$ centre-of-mass energy of 300--318 GeV for exchanged photon
virtuality, $Q^2$, above 1 $\gev^2$. The results support the
existence of a narrow baryonic state with strangeness in $\ksp$
and $\kspb$ decay channels, consistent with the pentaquark
prediction. No pentaquark signals were found in the $K^+ p$,  
$\xp$ and $\xbp$  channels.}

\maketitle

\section{Introduction}

This paper discusses recent ZEUS searches for
pentaquark baryons with strangeness in the $\ksppb$, $K^+p$,  
$\xp$ and $\xbp$ invariant-mass spectra in $ep$ collisions measured with the
ZEUS detector at HERA. The measurements were based on the central
pseudorapidity region,  $|\eta|\leq 1.5$,  where the contribution
from the fragmentation of the proton remnant is negligible. If a
baryon with five quarks is produced without the net baryon number
carried by the proton remnant, emerging from the emissions of
gluons and quarks in the hadronisation process, this would open a
new chapter in our understanding of non-perturbative QCD.

Even in case of production of the standard baryons, the absence of
a guiding principle of how to compose three quarks to form  a
baryon at the fragmentation stage leads to a few possible
baryon-production mechanisms ("diquark" and "popcorn"). In case of
baryons with five quarks, the situation is expected to be even
more mysterious. Alternatively, in $ep$ collisions, one may assume
a small contribution of the net baryon number to the
central fragmentation region. In this case, the pentaquark production is
expected to be not very different from fixed-target
experiments. However,
this effect, which should also be responsible for   
baryon-antibaryon production-rate asymmetry, 
is expected at the level of a few percents. 
Obviously, no antipentaquarks are expected for this
mechanism.

The pentaquark baryons  have been observed by a number of
fixed-target experiments in the $K^+ n$ decay channel
\cite{fixed}. According to the predictions of the chiral soliton
model \cite{zp:a359:305}, such a state can be interpreted as a
bound state of five quarks, i.e. as a pentaquark, $\Theta^+ =
uudd\bar{s}$. According to its quantum numbers, $\ksp$ and $\kspb$
decays are also possible. For the $\ksp$
channel, evidence for a corresponding signal has been found by
low-energy fixed-target experiments
\cite{ks}.
Recently, other strange pentaquarks 
($\Xi^{--}_{3/2}=dsds\bar{u}$ and $\Xi^{0}_{3/2}=dsus\bar{d}$)
were reported in $pp$ collisions by the NA49 experiment \cite{prl92:042003}. At
present, however,  this experiment  does  not confirm 
the $\Theta^+$ state, production rate of which is expected to
be higher.

%%%%%%%%%%%%%%%%%%%%%%%%%%%%%%%%%%%%%%%%%%%%%%%%%%%%%%%%%%%%%
\section{Evidence for a baryonic state decaying to $\ksppb$}
%%%%%%%%%%%%%%%%%%%%%%%%%%%%%%%%%%%%%%%%%%%%%%%%%%%%%%%%%%%%%

ZEUS has performed a search for pentaquarks in $\ksppb$  decay
channel \cite{zeus}. The analysis used deep inelastic scattering events
measured with exchanged-photon virtuality $Q^2\ge 1\gev^2$. The
data sample corresponded to an integrated luminosity of 121
pb$^{-1}$. The charged tracks were selected in the central
tracking (CTD) with $p_T\ge 0.15\gev$ and $|\eta|\le 1.75$,
restricting this study to a region where the CTD track acceptance
and resolution are high. 

The total number of $\ks$ with $p_{T}(\ks)>0.3\gev$ and $|\eta(\ks)|<1.5$,
identified using the decay mode $\ks\to\pi^{+}\pi^{-}$, 
was  867K. 
To eliminate contamination from $\Lambda (\bar{\Lambda})$ decays,
candidates with a proton mass hypothesis $M(p\> \pi)<1121\mev$
were rejected.

\begin{center}
\begin{minipage}[c]{0.48\textwidth}
\includegraphics[width=6.0cm,angle=0]{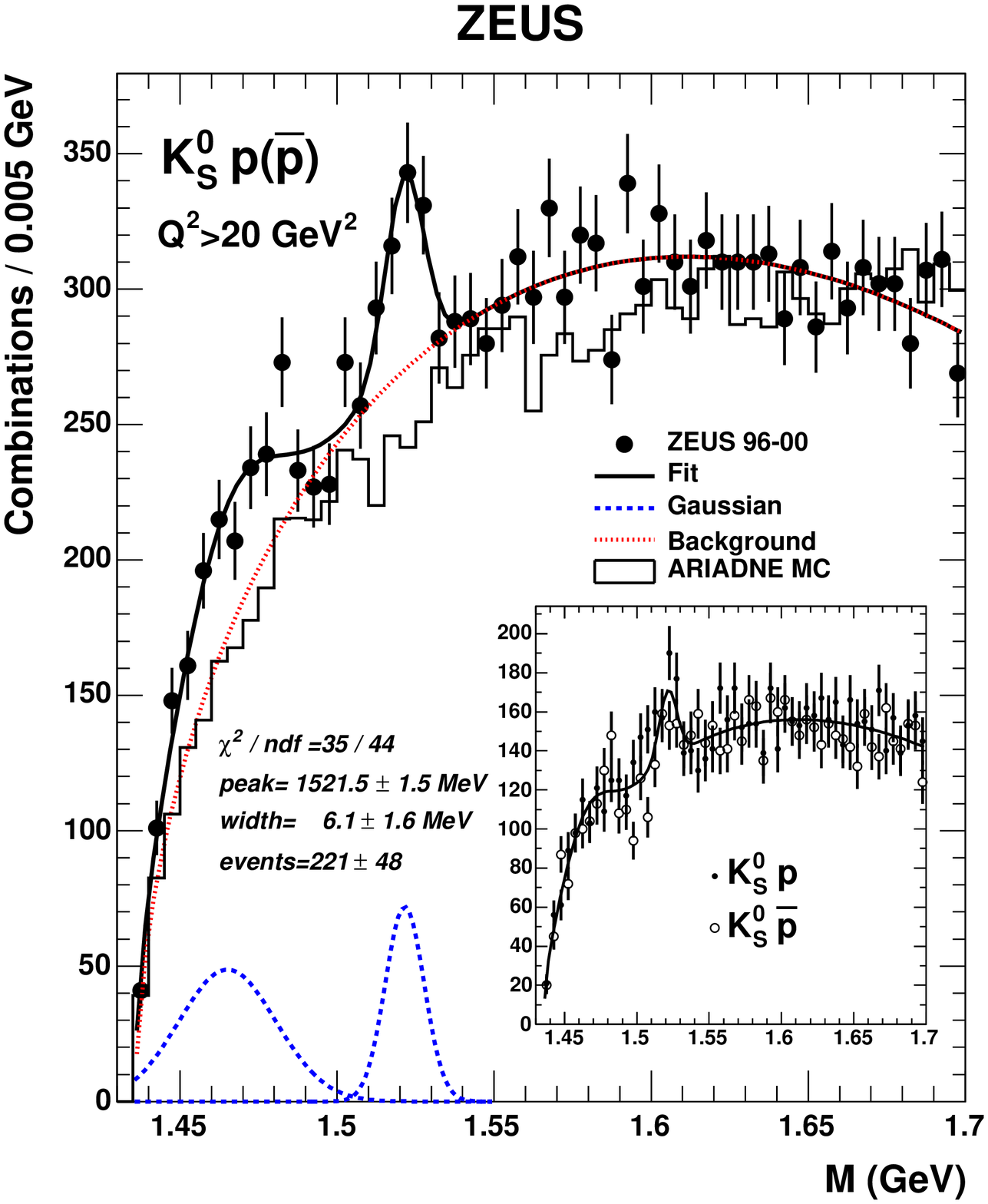}
\label{theta}
\end{minipage}
\hfill
\begin{minipage}[c]{.48\textwidth}
Figure 1: { Invariant-mass spectrum for the $\ksppb$ channel for
$Q^2 > 20 \gev^2$. The solid line is the result of a fit to the
data using the threshold background plus two Gaussians.  The
dashed lines show the Gaussian components, while the dotted line
indicates background. The prediction of the Monte Carlo
simulation is normalised to the data in the mass region above
$1650\mev$. The inset shows the $\kspb$ (open circles) and the
$\ksp$ (black dots) candidates separately, compared to the result
of the fit to the combined sample scaled by a factor of 0.5. }
\end{minipage}
\end{center}

The (anti)proton-candidate selection used the energy-loss
measurement in the CTD, $dE/dx$. $\ksppb$ invariant masses were
obtained by combining $\ks$ candidates in the mass region
$480-510\mev$ with (anti)proton candidates in the (anti)proton
$dE/dx$ band with the additional requirements $p<1.5\gev$ and
$dE/dx>1.15$ mips in order to reduce the pion background.  The CTD
resolution for the $\ksppb$ invariant-mass near $1530\mev$,
estimated using Monte Carlo simulations, was  $2.0\pm 0.5\mev$ for
both the $\ksp$ and the $\kspb$ channels.

Figure~1  shows the $\ksppb$ invariant mass  for $Q^2> 20\gev^2$,
as well as for the $\ksp$ and $\kspb$ samples separately (shown as
inset). 
The data is above the {\sc Ariadne} Monte Carlo model 
near  $1470\mev$ and 
$1522\mev$, with a clear peak at $1522\mev$. The signal extraction
was found to be difficult for very low $Q^2$ due to a large
background and acceptance effects.

To extract the signal seen at  $1522\mev$, the fit was performed
using a background function plus two Gaussians. The background has
the form
$F(M)=P_1(M-m_{K}-m_{p})^{P_2}(1+P_3(M-m_{K}-m_{p}))$, where $m_K$
and $m_p$ are the masses of the kaon and the proton, respectively,
and $P_{i=1,2,3}$ are free parameters. The first Gaussian, which
significantly improves the fit at low masses, may  correspond to
the unestablished PDG $\Sigma (1480)$. 
The peak position  
determined from the second Gaussian  was $1521.5\pm 1.5({\rm
stat.})^{+2.8}_{-1.7} ({\rm syst.}) \mev$. It agrees well with the
measurements by HERMES, SVD and COSY-TOF for the same decay
channel \cite{ks}.
If the width of the Gaussian is fixed to the
experimental resolution, the extracted Breit-Wigner width of the
signal was $\Gamma=8\pm 4(\mathrm{stat.}) \mev$.

The number of events ascribed to the signal by this fit was  $221
\pm 48$.  The statistical significance, estimated from the number
of events assigned to the signal by the fit, was  $4.6\sigma$. The
number of events in the $\kspb$ channel was $96 \pm 34$. It agrees
well with the signal extracted for the $\ksp$ decay mode.
If the observed signal corresponds to the pentaquark, this provides the first
evidence for its antiparticle with a quark
content of $\bar{u}\bar{u}\bar{d}\bar{d}s$.

The measured mass and the width of the observed state are close to
those observed in the $K^+n$ channel\footnote{
The systematical mass-scale uncertainties for  this 
decay mode are usually larger than for the $\ksppb$ channel.}, 
and agree well with the theoretical expectations
\cite{zp:a359:305}. The PDG reports no $\Sigma$ states in the
invariant-mass region 1480--1560 MeV, and no peak at a similar
mass was observed in $\Lambda\pi$ decays. This favours the
pentaquark explanation of the signal.

This is the first observation of  such  resonance
in high-energy colliding experiments.
Since the signal was observed in the central rapidity region,
and since the number of reconstructed particles agrees with the 
the number of reconstructed antiparticles, 
this indicates the fragmentation origin of the observed state.

%%%%%%%%%%%%%%%%%%%%%%%%%%%%%%%%%%%%%%%%%%
\section{$\Theta^{++}$ state?}
%%%%%%%%%%%%%%%%%%%%%%%%%%%%%%%%%%%%%%%%%%

If $\Theta^{+}$ is an isotensor state, a $\Theta^{++}$ signal can
be expected in the $K^+ p$ spectrum \cite{plb570:185}. The $K^\pm
p (K^\pm\bar{p})$ invariant mass spectra were investigated in  a
wide range of minimum $Q^2$ values, identifying proton and charged
kaon candidates using the $dE/dx$ information. The proton
candidates inside the $dE/dx$ proton band were required to have
$dE/dx>1.8$ mips, while the kaon candidates were reconstructed in
the kaon band after the restriction $dE/dx >1.2$ mips. For
$Q^2>1\gev^2$, no peak was observed near 1522 MeV in the $K^+p$
and $K^-\bar{p}$ spectra, see Fig.~2, while a clean signal was seen in
the $K^-p (K^+\bar{p})$ channel at $1518.5\pm
0.6(\mathrm{stat}.)\mev$, corresponding to the PDG
$\Lambda(1520)D_{03}$ state.  
No pentaquark was observed for $Q^2>20\gev^2$, where the signal
in the $\ksppb$ channel was best seen.

\begin{center}
\begin{minipage}[c]{0.58\textwidth}
\includegraphics[width=6.5cm,angle=0]{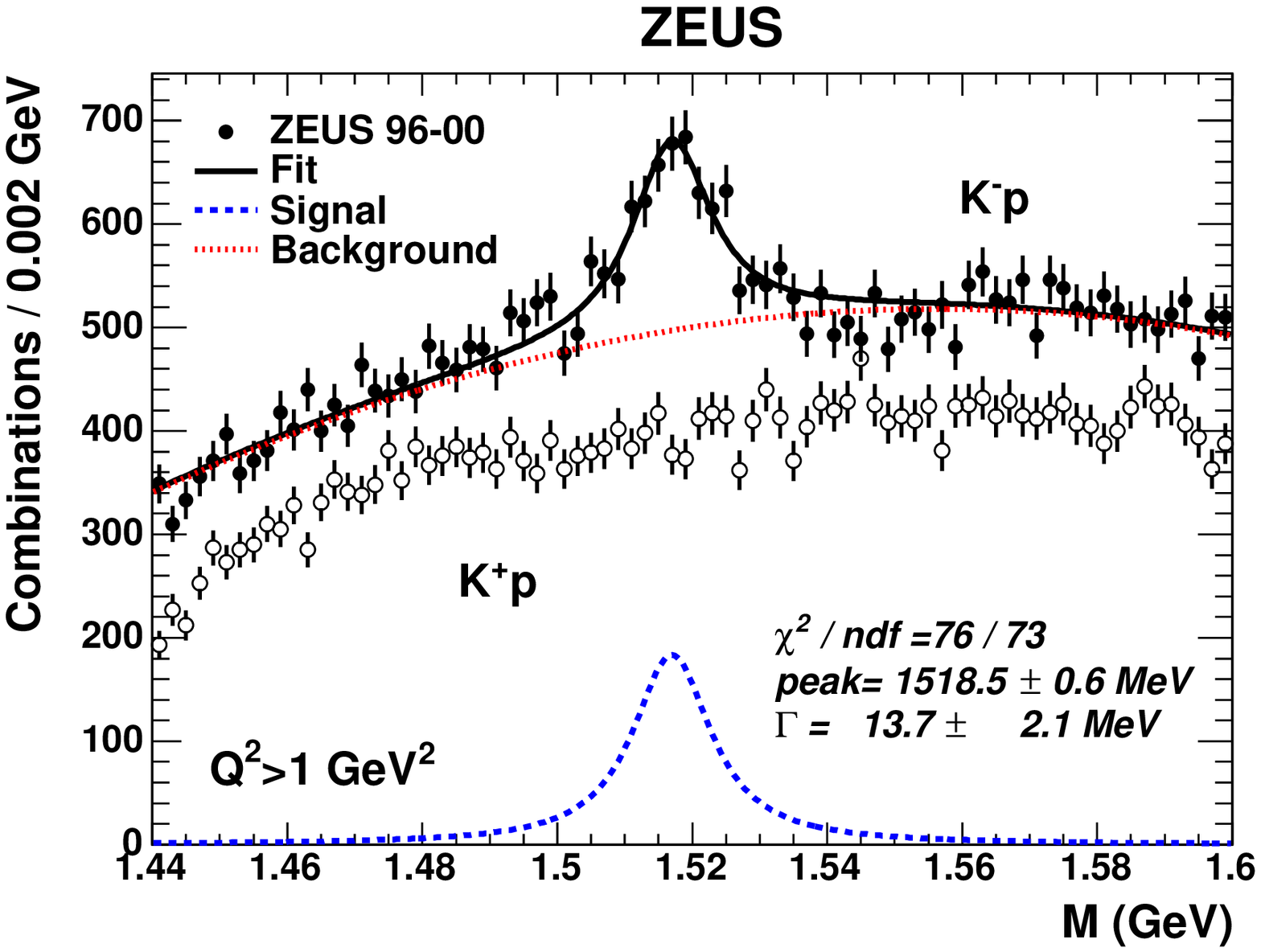}
\label{theta++}
\end{minipage}
%
% \hfill
\begin{minipage}[c]{.40\textwidth}
Figure 2: { Invariant-mass spectra for the $K^+p$ and $K^-p$
channels (plus charge conjugates) for $Q^2 > 1 \gev^2$. }
\end{minipage}
\end{center}

\begin{center}
\begin{minipage}[c]{0.58\textwidth}
\includegraphics[width=7.3cm,angle=0]{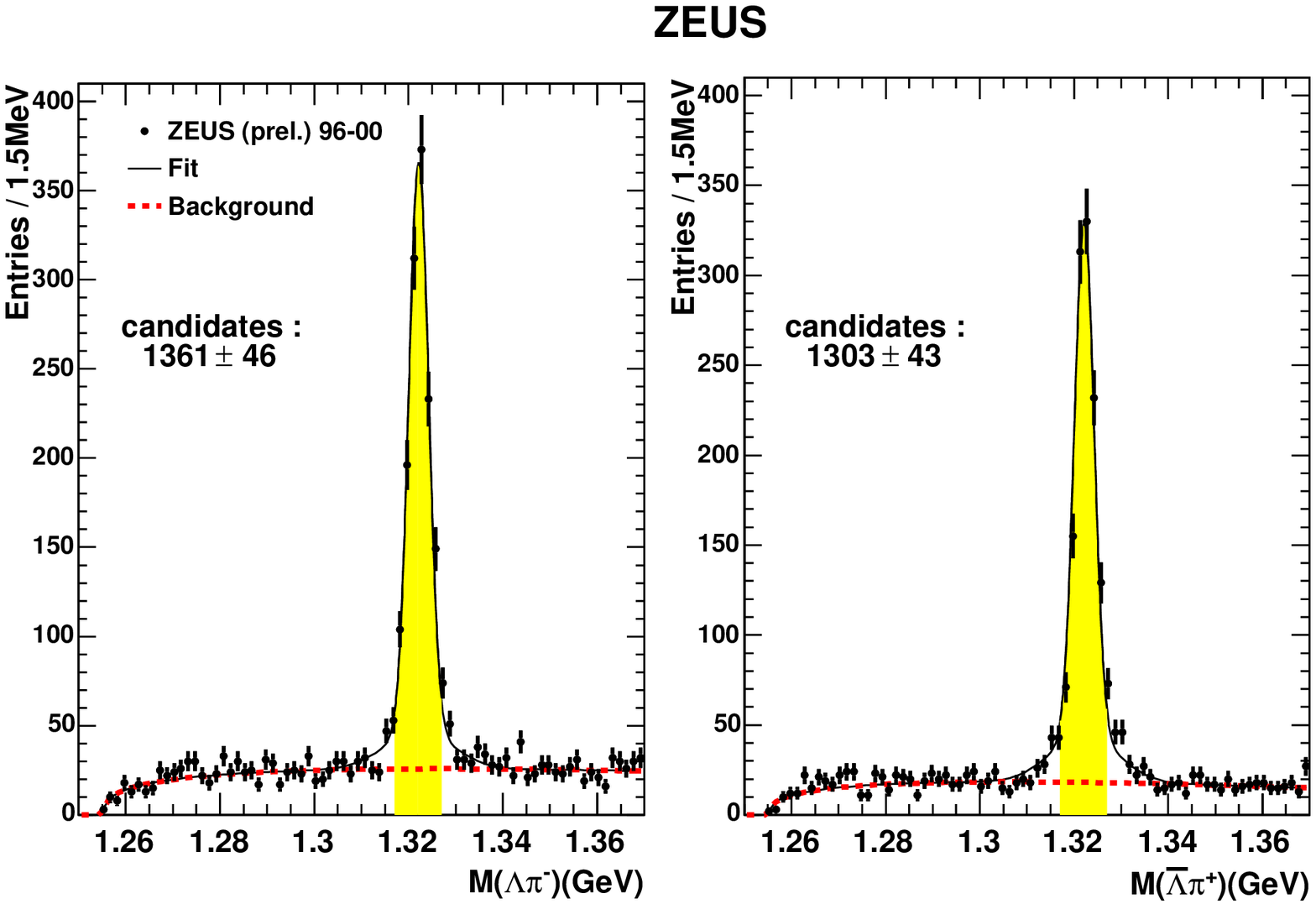}
% \label{theta++}
\end{minipage}
%
% \hfill
\begin{minipage}[c]{.40\textwidth}
Figure 3: {$\Lambda \> \pi$  invariant-mass spectra for
the $\Xi^-$ and $\bar{\Xi}^+$ states for $Q^2 > 1
\gev^2$. }
\end{minipage}
\end{center}

The failure to observe $\Theta^{++}$ indicates that the $\Theta^+$
state is not isovector or isotensor. Note that direct comparisons
of the production rates of possible pentaquarks and
$\Lambda(1520)$ states are not possible before taking into account
the detector acceptance effects.

%%%%%%%%%%%%%%%%%%%%%%%%%%%%%%%%%%%%%%%%%%
\section{$\Xi^{--}_{3/2}$  and $\Xi^{0}_{3/2}$ states?}
%%%%%%%%%%%%%%%%%%%%%%%%%%%%%%%%%%%%%%%%%%

The $\Theta^+$ lies at the apex of a hypothetical anti-decuplet of pentaquarks 
with spin $1/2$ \cite{zp:a359:305}. The baryonic states
$\Xi^{--}_{3/2}$ and $\Xi^{0}_{3/2}$ at the bottom of this
antidecuplet  are also manifestly exotic. Strong support for this
picture came from the NA49 experiment which recently made
observations of both states near $1862\mev$ in the 
$\xp$ and $\xbp$ decay channels in $pp$
collisions at the CERN SPS.

ZEUS has performed a similar analysis.  
First, $\Lambda$ candidates were reconstructed using a 
selection similar to that used for the $\ks$ reconstruction. The $\Xi$ signals, 
shown in Fig.~3,  were  reconstructed by combining $\Lambda$ with $\pi$
using displaced tertiary vertices \cite{andy}.

Figure~4 shows the $\Xi\>\pi$ invariant masses for  
four separate channels at
$Q^2>1\gev^2$, while Fig.~5 illustrates the combined sample.    
No pentaquark signal was observed near the $1860\mev$ mass region.    
A similar analysis was
performed for $Q^2>20\gev^2$, a kinematic region where $\Theta^+$ was best seen.  
However, again, no NA49 signal was found.

\begin{center}
\includegraphics[width=9.5cm,angle=0]{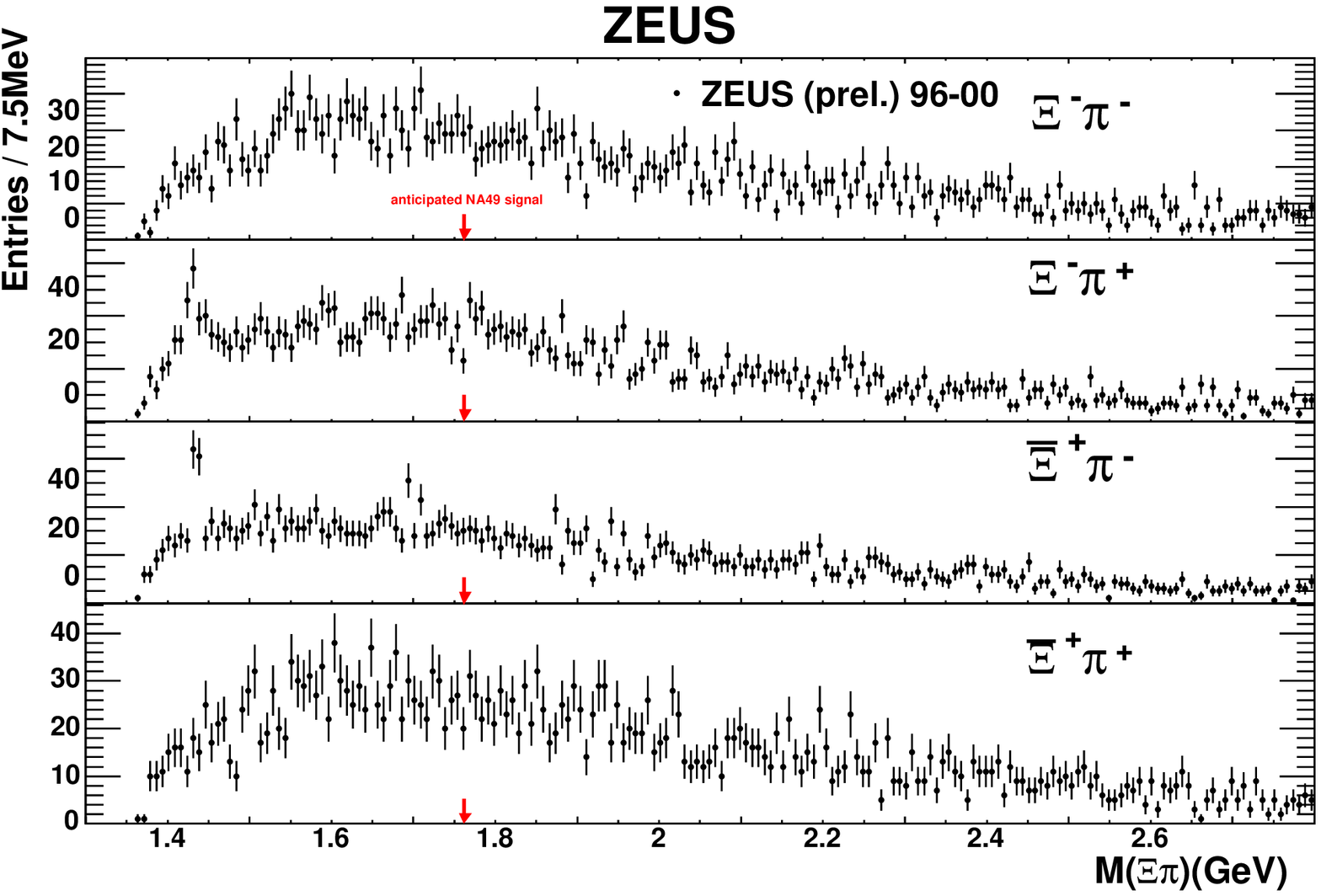}
\label{theta++0}
\end{center}
Figure 4:{
Invariant-mass spectra for
separate $\Xi^-\pi^-$, $\Xi^-\pi^+$,
$\bar{\Xi}^+\pi^-$ and $\bar{\Xi}^+\pi^+$ decay channels for
$Q^2> 1 \gev^2$. \vspace{0.5cm}}

In this decay channel, ZEUS clearly observes the $\Xi^*(1530)$
state, which can be used as gauge for the comparison of the
sensitivity to the pentaquark signal. With more than 160
$\Xi^*(1530)$ candidates reconstructed, the statistical sensitivity
exceeds the one of the NA49 measurement \cite{prl92:042003}, thus
the pentaquark signal should be difficult to miss. However, it should be
noted that NA49 is a fixed target experiment, which has good
acceptance in the forward region. Therefore, the absence of the
signal in ZEUS data may indicate that ZEUS has a little
sensitivity to this state, which may predominantly be produced in the
forward region. Interestingly, NA49 collaboration does not see the
$\Theta^+$ state, the production mechanism of which is unlikely to be
very different from that of the $\Xi^{--}_{3/2}$ and $\Xi^{0}_{3/2}$ states. Clearly,
this contradiction will need to be solved using future
high-luminosity HERA data and results from other experiments.

\begin{center}
\begin{minipage}[c]{0.65\textwidth}
\includegraphics[width=8.5cm,angle=0]{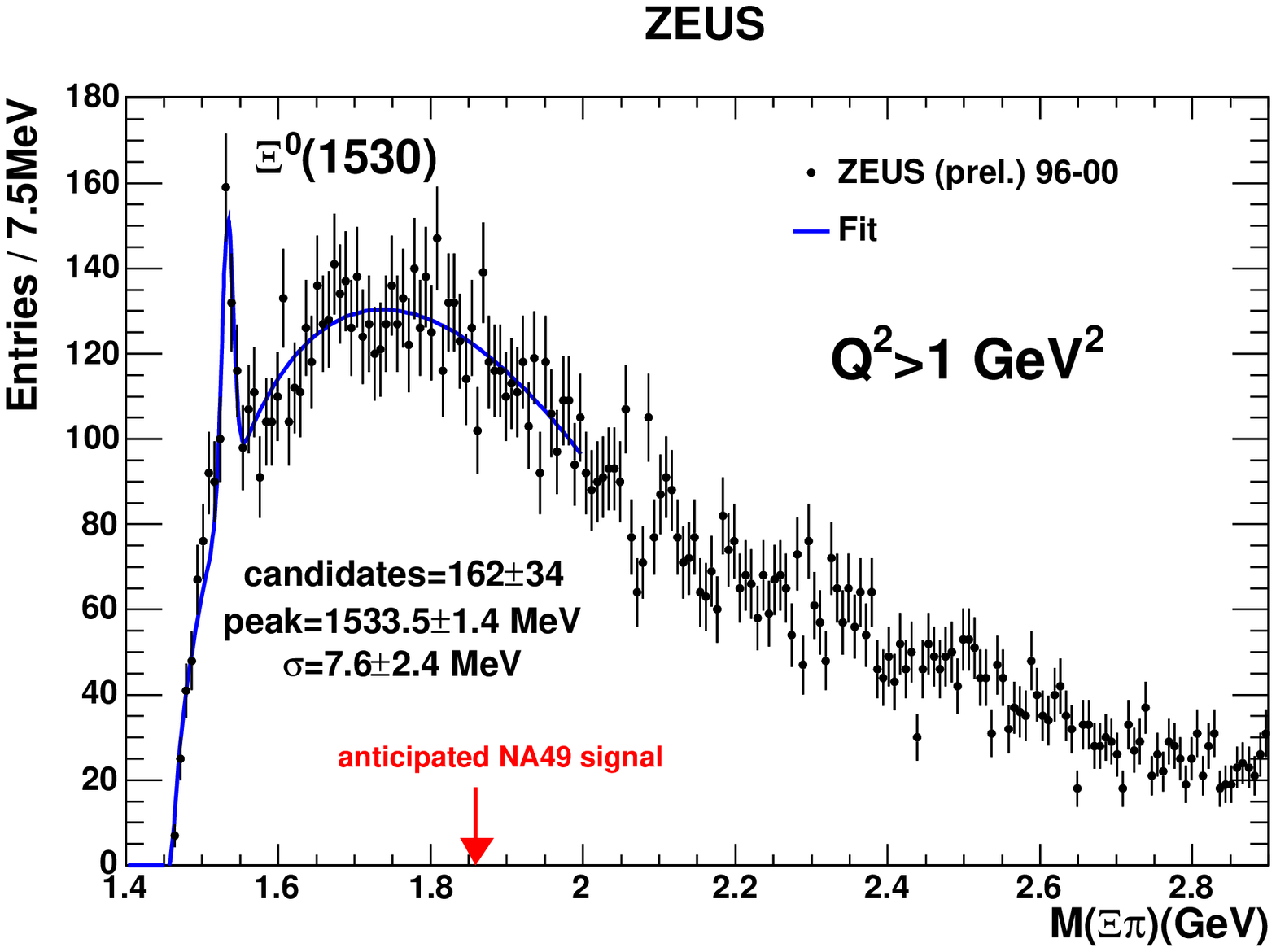}
\label{theta1++}
\end{minipage}
%
% \hfill
\begin{minipage}[c]{.32\textwidth}
Figure 5: {
Invariant-mass spectra for sum of the  $\xp$ and $\xbp$
channels for  
$Q^2> 1 \gev^2$.} 
\end{minipage}
\end{center}

\end{document}